\documentclass[final,12pt]{elsarticle}

\usepackage{amssymb,amsmath}

\usepackage{color}
\usepackage{lineno}
\usepackage{eurosym}
\usepackage{hyperref}
\journal{Journal of Optics}

\begin{document}

\begin{frontmatter}

\title{3D scattered light imaging: extracting 3D fiber orientations from 1D line profiles in brain imaging}

\author[inst1]{Dennis Scheidt}\ead{d.scheidt@fz-juelich.de}
\author[inst2]{Charlotte Voß}
\author[inst3]{Cristian Rosero Arias}
\author[inst3]{Santiago Saavedra Castano}
\author[inst1]{Felix Matuschke}
\author[inst4,inst1]{Roxana Koojimans}
\author[inst1,inst5]{Katrin Amunts}
\author[inst3]{Arturo Susarrey-Arce}
\author[inst1,inst2]{Markus Axer}


\affiliation[inst1]{organization={Institute for Neuroscience and Medicine 1, Forschungszentrum Juelich},
            addressline={Wilhelm-Johnen-Straße}, 
            city={Juelich},
            postcode={52428}, 
            country={Germany}}
\affiliation[inst2]{organization={School of Mathematics and Natural Sciences, University of Wuppertal},
            addressline={Gaußstraße}, 
            city={Wuppertal},
            postcode={42219}, 
            country={Germany}}
\affiliation[inst3]{organization={Department of Chemical Engineering, Mesoscale Chemical Systems, MESA+ Institute, University of Twente},
            addressline={PO Box 217}, 
            city={Enschede},
            postcode={7500 AE}, 
            country={The Netherlands}}
\affiliation[inst4]{organization={Netherlands Institute for Neuroscience, Royal Netherlands Academy of Arts and Sciences},
            addressline={Meibergdreef 47}, 
            city={Amsterdam},
            postcode={1105 BA}, 
            country={The Neatherlands}}
\affiliation[inst5]{organization={C. \& O. Institute of Brain Research, Heinrich-Heine University Düsseldor},
            addressline={Merowinger Platz 1a}, 
            city={Düsseldorf},
            postcode={40225}, 
            country={Germany}}

\begin{abstract}
\noindent 
Understanding the 3D fiber architecture of the brain at the microscopic scale is essential for revealing its structural connectivity and function. Polarization-based optical imaging (3D-PLI) techniques enable high-fidelity reconstruction of single nerve fiber orientations but struggle to resolve fiber crossings, which are critical for recovering the full connectome. Scattering-based imaging provides access to the structure factor of three-dimensionally oriented fibers. By probing a fixed scattering angle under multiple azimuthal illumination angles, in-plane fiber orientations and crossings can be recovered using computational scattered light imaging (SLI). However, despite containing 3D information, a theoretical framework to extract full 3D orientations has been lacking. In this talk, a simple analogical approximation of Rayleigh-Gans scattering theory is introduced to extract the 3D orientation of nerve fibers from one-dimensional SLI measurements. The theory is validated using tilted microscopic glass phantoms consisting of 2 µm-thick rod lattices fabricated by two-photon lithography. Finally, the method is applied to brain tissue samples and compared with 3D-PLI.
\end{abstract}

\begin{keyword}
Scattered Light Imaging \sep Brain Imaging \sep 3D Imaging \sep Two Photon Lithography \sep Tissue Phantom \sep Polarized Light Imaging
\PACS 0042 
\MSC 0078 
\end{keyword}

\end{frontmatter}

\section{Introduction}

Mammalian brains consist of millions to billions of neurons that are interconnected through synapses and form highly complex structural and functional networks \cite{connectome}. Understanding brain function and behavior therefore requires a multiscale perspective, ranging from molecular and cellular organization to mesoscopic fiber architecture and macroscopic connectivity \cite{connectome,Multiscale_brain}.

While the cellular composition and many anatomical regions of the brain have been extensively characterized, the detailed reconstruction of neuronal connectivity remains challenging \cite{brain_mapping}. In particular, mapping the trajectories and crossing patterns of nerve fibers at microscopic and mesoscopic length scales is still a major limitation. Diffusion tensor imaging (DTI) provides access to fiber orientations at the macroscopic scale, but its spatial resolution is insufficient to resolve microscopic fiber architecture \cite{DTI_brain_resolution}. Optical methods such as 3D-polarized light imaging (3D-PLI) and polarization-sensitive optical coherence tomography (PS-OCT) can reconstruct in-plane and three-dimensional nerve fiber orientations at microscopic resolution \cite{PLIpaper,PS_OCT_brain}. These techniques exploit the birefringence of the myelin sheath surrounding nerve fibers. 3D-PLI has also been successfully applied to cortical tissue, where fiber densities and birefringence signals are considerably lower than in white matter \cite{PLIpaper}.

However, birefringence-based methods have an intrinsic limitation in regions containing multiple crossing fiber populations. Since PLI measures the accumulated retardation of the transmitted light field, the resulting signal represents an effective orientation rather than the individual orientations of all contributing fiber populations. Consequently, two crossing fiber bundles with different in-plane orientations can yield a circular average of the underlying orientations, which prevents a direct separation of crossing fibers \cite{PLI_signal_no_crossings}.

In contrast, Scattered Light Imaging (SLI) exploits the anisotropic scattering properties of nerve fibers. Nerve fibers can be approximated as cylindrical structures that scatter light predominantly perpendicular to their symmetry axis. In SLI, the angular scattering response is measured by rotating an oblique illumination direction azimuthally around the sample \cite{SLIpaper}. A single in-plane fiber orientation then gives rise to two characteristic scattering peaks in the angular profile \cite{SLIpaper}. Since the measured intensity is approximately a linear superposition of the scattering contributions within one pixel, crossing fiber populations can generate multiple distinct scattering peaks. For example, two crossing in-plane fiber orientations can produce four peaks, allowing the individual orientations to be separated \cite{SLIpaper}.

Previous simulations and experiments indicate that the angular scattering profile also contains information about the out-of-plane inclination of nerve fibers \cite{SLIpaper}. However, a closed analytical description that directly relates the measured peak geometry to the three-dimensional fiber orientation has so far been missing. In particular, a model is required that links the angular peak positions and peak separations to the underlying in-plane orientation and inclination of cylindrical scatterers.

In this work, we introduce an analytical model based on the Rayleigh-Gans approximation for cylindrical scatterers to recover the missing inclination information from SLI measurements. This extends SLI from an in-plane orientation imaging technique toward a three-dimensional fiber orientation method, analogous in scope to 3D-PLI but with the additional ability to resolve crossing fiber configurations. 
The ground-truth orientation of brain fibers is generally not known \textit{a priori} because of tissue complexity and inter-subject variability. Therefore, microscopic glass phantoms containing crossing cylindrical structures were fabricated using two-photon lithography to systematically investigate the relationship between fiber architecture and the measured SLI signal \cite{tpl_glass}.
The glass cylinders have diameters of $1-2$ \textmu m to approximate the diameter range of large axons \cite{axon_diameter}.
These phantoms were measured under different tilting angles to emulate inclined nerve fibers and to validate the proposed analytical reconstruction method. Finally, the model is applied to a human brain section, and the reconstructed inclination is compared with inclination estimates obtained by 3D-PLI.

The results demonstrate that the Rayleigh-Gans model of a transparent cylinder provides a suitable first-order approximation for describing the SLI signal in both glass phantoms and human brain tissue. Beyond the in-plane orientation of crossing fibers, the angular scattering profiles also contain information about the effective fiber diameter and inclination. In white matter, the reconstructed fiber diameters are substantially larger than in gray matter, consistent with the expected anatomical composition. The inclination measurements show a strong correlation with 3D-PLI in white matter for low inclination angles. Remaining deviations may result from the different contrast mechanisms and angular sensitivities of the two imaging techniques and will be investigated in future work.

\section{Methods}
\subsection{Setup}

\begin{figure}
    \centering
    \includegraphics[width=0.95\linewidth]{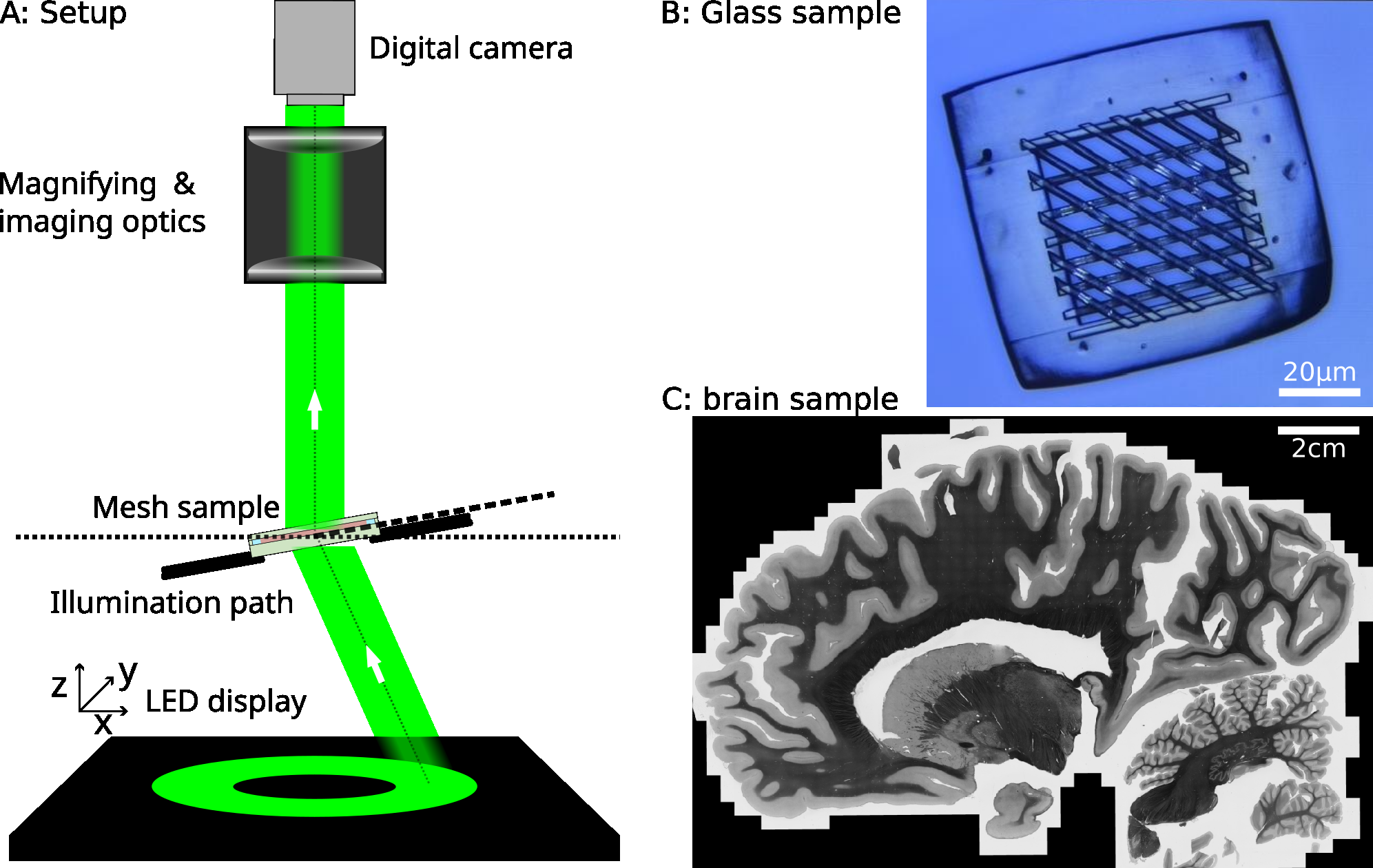}
    \caption{Experimental setup and used samples. A: Experimental setup with LED display, sample stage, imaging lens and camera. The stage is rotated to simulate inclinated samples. B: Microscopic glass phantom created with two photon lithography. C: Sagittal human brain section.}
    \label{fig:setup}
\end{figure}

The measurements were performed with the setup illustrated in Fig.\ref{fig:setup}. The optical system follows the general ComSLI configuration reported previously in \cite{Francas_Article,Scheidt:26_Hadamard} and can also be operated for Mueller-polarimetric measurements. In the present work, however, only the scattered-light imaging modality was used. Therefore, the polarimetric elements were positioned outside the relevant scattered-light detection path and are omitted from the schematic for clarity.

Oblique illumination was provided by a $46.56 \times 46.56$ cm$^2$ LED display (INFILED s1.8 LE Indoor LED Cabinet, frame rate 2400 Hz). The display contains $256 \times 256$ individually addressable RGB LEDs with a pitch of $1.8$ mm and a viewing angle of $120^\circ$. The sample was mounted 16 cm above the display, which corresponds to the focal distance of the imaging objective. Annular illumination patterns were displayed with an inner radius of 50 LEDs and an outer radius of 90 LEDs, resulting in an approximate polar illumination angle of $45^\circ$.

Image acquisition was performed with a Basler acA5472-17uc camera equipped with a Sony IMX183 CMOS sensor. The sensor has $5472 \times 3648$ pixels with a pixel size of $2.4 \times 2.4$ \textmu m$^2$. A QIOPTIQ APO-RODAGON-D 1X 75/4.0 objective was used for imaging, yielding a field of view of approximately $8 \times 5$ mm$^2$ and an effective object-space pixel size of about $1.5$ \textmu m. The optical resolution was estimated to be $2.19$ \textmu m using a USAF resolution target. To maximize the collected scattered-light signal, the camera aperture was fully opened to a diameter of 5.6 mm.

For all measurements, only the green LEDs of the display were used, enabling comparability with previous ComSLI experiments performed under green illumination \cite{Francas_Article}. The exposure time was set to 0.5 s for the Hadamard sampling scheme described in \cite{Scheidt:26_Hadamard}. Camera gain was not applied, and the analysis was restricted to the green channel of the recorded images.

For tilted measurements, the sample stage was placed on a optical post mount and fixated at the vertical axis. Then the stage was rotated in steps of $\Delta \epsilon = 5^\circ$ from $\epsilon =$ [$0^\circ,70^\circ$]. 
Tilt positions were verified with a digital lever (Stabila TECH 196 DL). After $\epsilon = 70^\circ$ no further measurements were possible, as the stage then blocks the illumination part partially, so that the sample is not illuminated from all sides anymore.
A brain tissue sample was measured in a Polarization Microscope LMP3D (Taorad GmbH, Aachen, Germany) \cite{taorad}. There, first the PLI measurement with 5 different tilts of the illumination path were conducted. Afterwards, a customized LED ring with $N=32$ LEDs performed the SLI measurement. The whole section was scanned and stitched using an in-house stitching algorithm.  

\subsection{Sample preparation}

Two samples were investigated. First, a glass phantom fabricated by two-photon lithography was used to validate the scattered-light signal model:\\
A custom made silicon loaded photoresin formulation was prepared under yellow light conditions to reduce UV exposure and prevent undesired polymerization. POSS and PETA were mixed in a rotary evaporator operated under reduced pressure at 75 °C, and 140 rpm. After having those components homogeneously mixed, DETC was dissolved in DMAc with a vortex mixer and then transferred to the POSS and PETA mixture. The mixing continued for another hour, but under ambient pressure. The final yellow-colored photoresin was pipetted into 25 mL amber bottles (Duran) and used directly to obtain the pre-glass mesh microarchitecture employing a Two Photon Litography (TPL) direct laser writing system (Photonic Professional GT, Nanoscribe) operated in the DiLL mode. A 25×/0.8 objective (Imm Corr DIC M27, LCI Plan-Neofluar; Carl Zeiss) focused the femtosecond laser radiation focal spot on the resin using differential interference contrast. For this printing, an ITO-coated glass slide (2.4 cm × 3.2 cm) was used as a substrate to keep the microarchitectures in place. Finally, the TPL printed microstructure, together with the substrates, were placed in open-cap ceramic crucibles and then transferred into an air chamber oven (LH 15/12, Nabertherm). The thermal processing was conducted at 650 °C to combust the organic constituents from the composite polymer and transform the 3D mesh into the desired glass. More details on the fabrication process can be found in \cite{tpl_glass}.
The microstructure consists of nearly cylindrical glass rods with a diameter of $2 \mu\mathrm{m}$, arranged at a predefined crossing angle of $45^\circ$ (see Fig. \ref{fig:setup} B). To reduce diffraction artifacts caused by periodic grating-like arrangements, the rod positions were randomized using Poisson-disk sampling with a minimal distance of $5\mu$m. The rods were embedded in a rectangular glass frame with a thickness of $20\mu\mathrm{m}$, which provided mechanical stability during the cooling-induced shrinkage process after fabrication.

Second, a sagittal section of a human brain was investigated.
The post-mortem brain sample (female, 80 years old) was formalin-fixed and deep-frozen as described above. The frozen brain was cut into 60\,\textmu m-thin sections which were mounted in a $20\%$ glycerol-water solution in between coverslips. One section was selected for this study (see Fig. \ref{fig:setup} C). Detailed preparation protocols are described in \cite{Tissue_prep}.

This study was conducted in accordance with all applicable ethical regulations. The body donor provided written informed consent for the general use of post-mortem tissue for research and educational purposes.
The use of the brain was approved by the Netherlands Brain Bank (ethics approval NBB-1037/2018), in the Netherlands Institute of Neuroscience, Amsterdam, the Netherlands.

\subsection{Signal model of a cylindrical scatterer}\label{sec:model}

The diameter of myelinated axons is distributed approximately in the range of $d = [0.5 - 10]\mu$m, with an average value of $d_{avg} \approx 1\mu$m \cite{axon_diameter}. This is on the order of the illumination wavelength $\lambda = 530$nm. Consequently, the scattering process cannot, in general, be described by simple geometric optics. Instead, the interaction of light with axons has to account for wave-optical effects such as interference, multiple internal reflections, and scattering from finite-sized dielectric structures \cite{scattering_book}. A full description would require Mie-type scattering theory. However, for finite cylindrical structures and the specific illumination--detection geometry used here, closed analytical solutions are not available in a form suitable for large-scale image analysis. Numerical solutions of the corresponding scattering problem are computationally expensive and therefore inefficient for pixelwise reconstruction in large microscopy datasets.

A tractable approximation is provided by the Rayleigh-Gans (RGS) theory, in which the scattering object is approximated as a continuous distribution of weakly scattering dipoles. This approximation is valid under the condition $kd|\Delta n| \ll 1$, where $k = 2\pi/\lambda$ is the wave number and $\Delta n = n_m - n_s$ is the refractive-index difference between the surrounding medium $n_m$ and the sample $n_s$. In brain tissue, typical refractive indices are in the range $n_s \approx 1.36 - 1.42$, while the mounting medium used here, a 20\% glycerol-water solution, has approximately $n_m = 1.36$. Thus, $|\Delta n| \approx 0.01 - 0.06 \ll 1$. Although the Rayleigh-Gans condition is most strictly fulfilled for smaller axon diameters and weak refractive-index contrasts, the approximation provides an analytically accessible model that captures the dominant angular structure of the scattering pattern.

For unpolarized light, which is used in the experiments, the scattered intensity $I_{scat}$ of a cylindrical scatterer can be expressed as
\begin{equation} \label{eq:I_scat}
   I_{scat}(\theta,\phi) =  \frac{1+\cos^2(\theta_{scat})}{2} \frac{(kr)^4h^2}{4R^2}|\Delta n| I_0 D^2(\theta_{scat},\gamma) \ ,
\end{equation}
where $\theta_{scat}$ is the scattering angle, $r$ is the cylinder radius, $h$ is the cylinder height, $R$ is the distance between the scatterer and the detector, $I_0$ is the incident intensity, and $D(\theta_{scat},\gamma)$ is the form factor of the scatterer. The form factor is obtained by integrating over the circular cross-section of the cylinder:
\[
D(\theta_{scat},\gamma) =
\frac{1}{\pi r^2}
\int_{-r}^{r}
\exp\left(2ik \sin(\theta_{scat}/2) y \sin \gamma\right)
2 \sqrt{r^2-y^2} \, dy ,
\]
which yields \cite{scattering_book}
\begin{equation} \label{eq:form_factor}
    D(\theta_{scat},\gamma) = \mathrm{sinc}\left( k h \sin (\theta_{scat}/2) \cos \gamma \right) \frac{ \mathrm{J}_1( 2kr\sin(\theta_{scat}/2) \sin \gamma )}{kr\sin(\theta_{scat}/2) \sin \gamma} \ .
\end{equation}
In this expression, the sinc term describes the axial contribution of the cylindrical scatterer, while the Bessel term $\mathrm{J}_1(r)/r$ describes the radial contribution. The remaining geometrical quantities, namely the scattering angle $\theta_{scat}$ and the effective angle $\gamma$, can be derived by describing both the incident light direction and the cylinder orientation in spherical coordinates. Here, $\alpha$ and $\beta$ denote the azimuthal and polar angle of the incident light, while $\phi_{cyl}$ and $\theta_{cyl}$ denote the azimuthal and polar orientation of the cylinder.

For the specific imaging geometry of the setup shown in Fig. \ref{fig:setup}A, the scattering angle is given by $\Theta_{scat} = \beta$. The effective angle $\gamma$ between the cylinder axis and the scattering vector is then given by \cite{scattering_book}
\begin{equation}\label{eq:scattering_angle}
    \gamma = \arccos\left(- \cos\theta_{cyl} \sin(\beta/2) + \sin(\theta_{cyl}) \cos(\beta/2) \cos(\alpha 
    + \phi_{cyl})\right) \ .
\end{equation}

\begin{figure}
    \centering
    \includegraphics[width=0.95\linewidth]{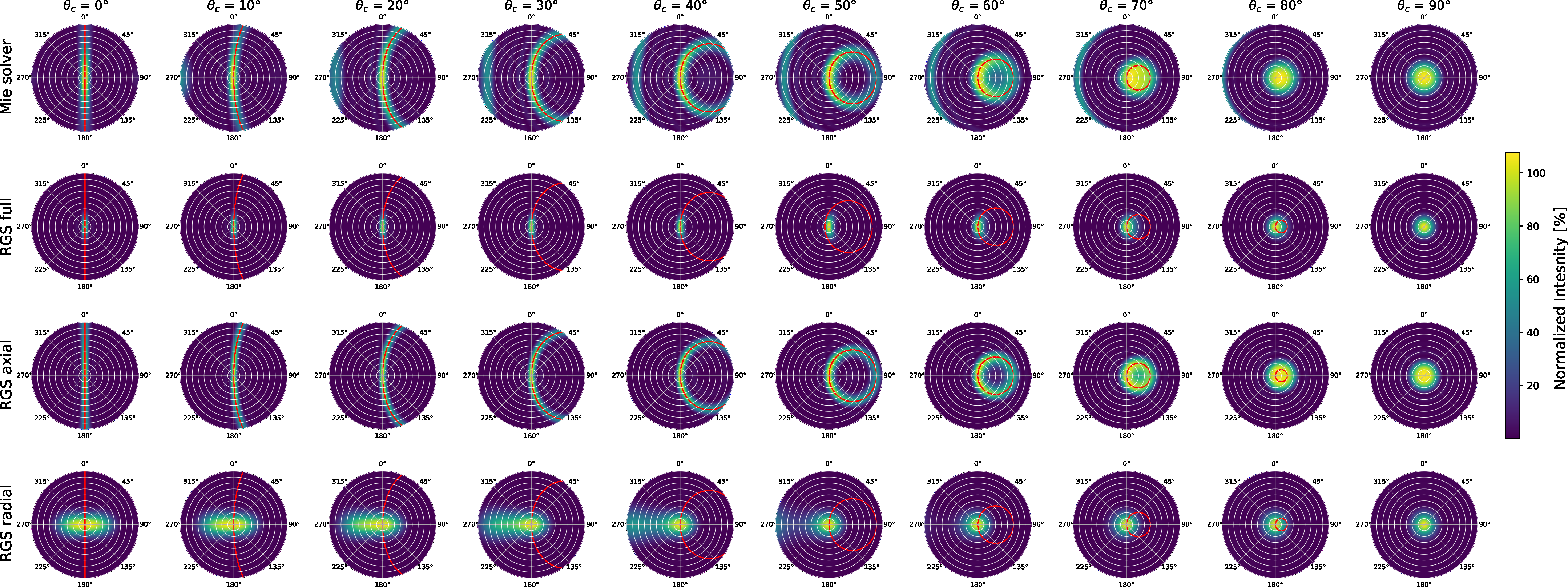}
    \caption{Scattering patterns of a single cylinder with radius $r = 500,\mathrm{nm}$ and height $H = 1\mu\mathrm{m}$ for the illumination geometry shown in Fig.~\ref{fig:setup}A. The cylinder inclination $\theta_{\mathrm{cyl}}$ increases from $0^\circ$ on the left to $90^\circ$ on the right in steps of $10^\circ$. The first row shows the numerical solution obtained with the ADDA solver~\cite{adda_YURKIN20112234}. The second row shows the corresponding result calculated from the full Rayleigh--Gans scattering (RGS) expression in Eq.~\ref{eq:form_factor}. The third and fourth rows show the isolated contributions of the axial $\mathrm{sinc}^2$ term and the radial $\mathrm{bsinc}_1^2$ term, respectively. The angular coordinate denotes the azimuthal illumination angle $\alpha$, whereas the radial coordinate denotes the polar illumination angle $\beta$. In the experiments, angular scattering profiles are sampled at a fixed polar angle $\beta$ while varying the azimuthal angle $\alpha$. }
    \label{fig:Mie_RGS}
\end{figure}

Figure \ref{fig:Mie_RGS} compares the scattered intensity of a cylinder with $r = 500$nm and $h = 1\mu$m for a numerical Mie-type solution obtained with the ADDA solver \cite{adda_YURKIN20112234} and for the RGS model described by Eqs. \ref{eq:I_scat}, \ref{eq:form_factor}, and \ref{eq:scattering_angle}. Different inclination angles $\theta_{cyl}$ are shown for a fixed in-plane orientation of $\phi_{cyl}= 45^\circ$. In the polar plots, the radial axis corresponds to the scattering angle $\Theta_{scat} = \beta$, while the polar axis corresponds to the illumination azimuth $\alpha$. The RGS solution is separated into its radial and axial contributions. The red lines indicate the expected positions of the scattering maxima according to conical diffraction theory \cite{conical_diffraction}. Experimentally measured angular scattering profiles correspond to samples taken at a fixed polar angle.

The comparison shows that the radial contribution of the RGS model mainly damps the signal with increasing polar angle and remains approximately constant for a fixed polar angle. Therefore, for the angular profiles measured at a fixed scattering angle, the dominant angular modulation can be described by the axial sinc$^2$ term. This becomes particularly clear when comparing the RGS model with the numerical Mie-type solution. The main angular structure of the scattering pattern is reproduced by the axial sinc$^2$ contribution, and the positions of the maxima agree well with the maxima predicted by conical diffraction theory. This indicates that the simplified analytical model is suitable for describing the peak positions in the measured angular scattering profiles.

With increasing inclination $\theta_{cyl}$, the scattering pattern changes from a nearly straight line to a curved structure. The ends of the curve shift toward the inclined side of the cylinder, in agreement with the observations reported by Menzel et al. \cite{SLI_simulation}. Consequently, if the scattering angle $\Theta_{scat}$, or equivalently $\beta$, is known, the inclination can be reconstructed from the angular peak separation $\Delta \alpha_s$ in the measured SLI profile. Solving the peak-position condition of the axial sinc$^2$ term yields
\begin{equation} \label{eq:inclination_peak_separation}
    \theta_{cyl}^{\mathrm{SLI}} = \arctan
    \left[
        \frac{
            \sin\left(
                \frac{\pi - \Delta \alpha_s}{2}
            \right)
        }{
            \tan\left(
                \frac{\beta}{2}
            \right)
        }
    \right].
\end{equation}

In the original RGS expression, the physical cylinder height $h$ appears in the argument of the axial sinc$^2$ term. For nerve fibers, however, the physical length can range from several micrometers to centimeters and is therefore much larger than both the fiber diameter and the optical resolution. Two interpretations are possible. First, the effective height could be limited by the camera pixel size and the optical magnification, such that the finite detection area defines the relevant scattering length. Second, for elongated fibers with $h/r \gg 1$, the fiber can be treated as effectively infinite along its axis. In this case, the finite transverse extent of the fiber becomes the dominant geometrical scale that determines the observed angular peak width. Under this effective-cylinder interpretation, the diameter of the cylinder or nerve fiber enters the sinc$^2$ term as the relevant structure size.

Under this assumption, the cylinder diameter can be estimated from the measured peak width. Solving the sinc$^2$ profile for the Full Width at Half Maximum (FWHM) yields:
\begin{equation} \label{eq:diameter_rec}
    D_{\mathrm{FWHM}}
    =
    \frac{
        2 \cdot 2.78 \cdot \pi
    }{
        \sqrt{2} \, k \, w_{\mathrm{FWHM}}
    }.
\end{equation}
Here, $w_{\mathrm{FWHM}}$ denotes the angular full width at half maximum of the scattering peak. Applying this relation to previous Hadamard-based experiments \cite{Scheidt:26_Hadamard} indicates that the observed peak width cannot be explained by the effective camera pixel size alone. In these experiments, strong peaks with a FWHM of approximately $3 \cdot 360/64^\circ$ were observed, corresponding to an estimated diameter of $D\approx 5\mu$m. This value is larger than the effective pixel resolution of $\Delta x = 1.5\mu$m and therefore supports the interpretation that the peak width contains information about the physical transverse structure size rather than being solely limited by the imaging pixel size.

\subsection{Peak-pair attribution algorithm}\label{sec:algorithm}

A single fiber population can generate up to two scattering peaks in the angular SLI profile. The angular separation of these peaks,
$\Delta \alpha_s \in [0,180]^\circ$, depends on the fiber inclination according to Eq.~\ref{eq:inclination_peak_separation}. For a single predominant fiber orientation within a local tissue region, the inverse problem is therefore comparatively well constrained: the in-plane orientation is determined by the angular position of the peak pair, while the inclination is inferred from the peak separation. However, when multiple fiber orientations are present within the same region, the individual scattering contributions are approximately linearly superposed. For $M$ different fiber orientations, up to $j=2M$ peaks may therefore be expected. Since peaks from different fiber populations can overlap, merge, or fall below the noise floor, the actually observed number of peaks can vary between $j=1$ and $j=2M$.

In practice, the attribution problem is limited to at most $M=3$ fiber orientations, corresponding to up to six detectable peaks. This restriction is motivated by both the finite angular sampling of the SLI profile and the expected biological complexity of the investigated tissue. For a profile with $N=32$ angular measurements, resolving and assigning more than six peaks would lead to a strongly underdetermined combinatorial problem. Moreover, crossing regions in white matter typically contain two dominant fiber orientations and only rarely require three \cite{Crossings_brain,SLIpaper}. In cortical myeloarchitecture, the dominant anisotropic scattering signal is also often described by a mesh-like arrangement of a small number of local orientations \cite{PALOMEROGALLAGHER2019716}. More complex configurations may occur in gray matter, but the reduced axon density in this region results in a weak anisotropic signal, such that reliable attribution of many individual peak pairs becomes increasingly noise-limited.

The measured angular profile $I(\alpha)$ is modeled as a linear superposition of peak-pair contributions:
\begin{equation}
I(\alpha) \approx B(\alpha) + \sum_{m=1}^{M} A_m , T_m(\alpha;\phi_{cyl,m},\Delta\alpha_{s,m},w_m),
\qquad M \leq 3,
\end{equation}
where $B(\alpha)$ denotes a slowly varying background, $A_m$ is the amplitude of the $m$-th contribution, and $T_m$ is a symmetric peak-pair template with orientation $\phi_{cyl,m}$, peak separation $\Delta\alpha_{s,m}$, and angular width $w_m$. The two peak positions of one candidate pair are given by:
\begin{equation}
\alpha_{m,\pm} =
\phi_m \pm \frac{\Delta\alpha_{s,m}}{2}
\quad \mathrm{mod}; 180^\circ .
\end{equation}
Equivalently, for two detected peak positions $\alpha_i$ and $\alpha_j$, the axial peak separation is computed as:
\begin{equation}
\Delta\alpha_{s,m}^{ij}
=
\min\left(
|\alpha_i-\alpha_j|,
180^\circ - |\alpha_i-\alpha_j|
\right),
\end{equation}
and the corresponding candidate orientation $\phi_{cyl,m}$ is estimated from the axial midpoint of the two peak positions. The inclination is then obtained by inserting $\Delta\alpha_{s,m}^{ij}$ into the inverse relation of Eq.~\ref{eq:inclination_peak_separation}.

The attribution procedure first detects local maxima in the angular profile and characterizes each peak by its position, prominence, and width. All geometrically admissible peak pairs are then generated from the detected peaks. Candidate pairs are rejected if their separation is outside the physically meaningful range or if they violate minimum angular-distance constraints. Each remaining pair receives a score based on peak prominence, approximate symmetry, peak width, template agreement, and consistency with the expected number of fiber populations. The best set of non-conflicting peak pairs is then selected by minimizing a combined attribution cost:
\begin{equation}
\mathcal{C}
=
\sum_{m=1}^{M}
\left[
\lambda_r , \mathrm{NRMSE}_m
-
\lambda_p , P_m
+
\lambda_s , S_m
+
\lambda_o , O_m
\right],
\end{equation}
where $\mathrm{NRMSE}_m$ measures the residual between the local profile and the peak-pair template, $P_m$ represents the combined peak prominence, $S_m$ penalizes asymmetric or implausible peak-pair geometry, and $O_m$ penalizes invalid overlap or duplicate assignments. The constants $\lambda_r$, $\lambda_p$, $\lambda_s$, and $\lambda_o$ weight the individual criteria.

Special care is required for overlapping peaks. If one observed maximum is shared by two candidate peak pairs, the shared peak is only accepted if it is sufficiently broad. This constraint is motivated by simulations showing that true peak superposition produces a widened peak, whereas isolated sharp maxima are more likely to represent individual peaks or noise-induced detections. Therefore, sharp but prominent peaks are allowed to form ordinary peak pairs, but they are not used as shared peaks between multiple orientations. Only peaks whose angular width exceeds a predefined threshold are considered valid shared-peak candidates. This prevents the attribution algorithm from explaining several fiber orientations by repeatedly reusing the same narrow peak.

The output of the peak-pair attribution step consists of initial estimates for the in-plane orientation, inclination, and diameter of each selected fiber population. 

\section{Results}

\subsection{Signal characterization at glass phantoms} \label{sec:characterization}
To characterize the scattering signal at intersecting structures, the crossing angles of the glass phantoms are determined using SLI and compared with the theoretically expected values. The glass phantoms used have predefined crossing angles of 90°, 60°, 45° and 30°, which also allows the influence of the crossing angle on the SLI signal to be investigated. As both glass rods and axons are described as transparent cylinders in the model used, the findings can, in principle, be extrapolated to brain tissue.

The glass phantoms are measured within the SLI setup and masked according to their averaged intensity, resulting in an individual mask for each phantom. In addition, the individual horizontal and vertical rods are manually masked, as this is required for determining their diameters.

\begin{figure}
    \centering
    \includegraphics[width=0.95\linewidth]{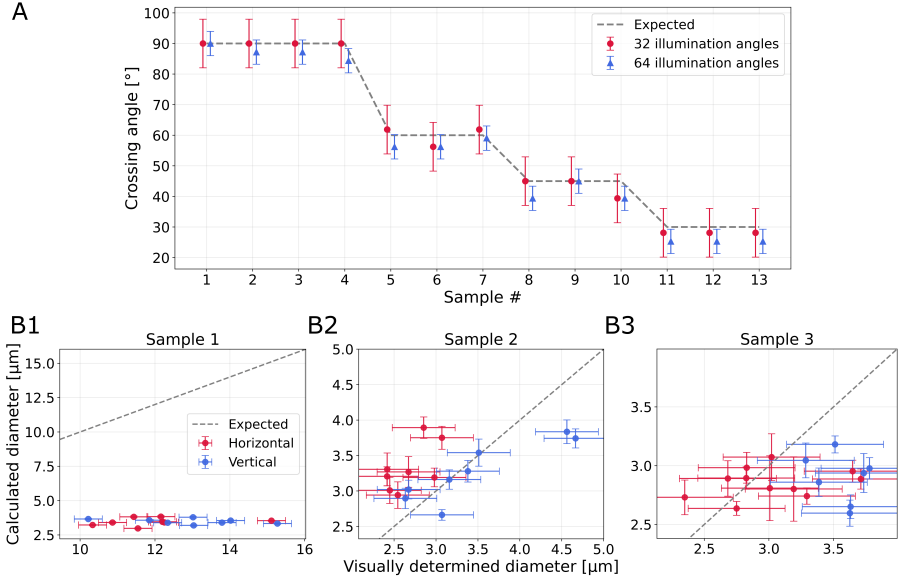}
    \caption{Parameter extraction of glass samples shown in Fig. \ref{fig:setup}B. A: Comparison of the reconstructed crossing angles obtained from SLI with the theoretical crossing angles of the glass phantoms for 32 (red) and 64 (blue) illumination directions. Error bars represent the standard deviation. B1-B3: Comparison of the visually determined and reconstructed diameters of the horizontal (red) and vertical (blue) rods for three glass phantom samples. The dashed line indicates the expected agreement between both values. Horizontal error bars represent the estimated uncertainty of the visual diameter determination, while vertical error bars represent the standard error of the mean (SEM) of the reconstructed diameter.}
    \label{fig:CrossingAngle_Diameter}
\end{figure}

To validate the analysis, the crossing angles are determined experimentally from the averaged scattering profiles. 
To do this, the intensity maxima are first detected and then assigned to pairs of peaks, as a glass rod produces two scattering peaks for two opposite illumination directions. The crossing angle is then calculated from the orientation of the peak pairs. Averaging the scattering profiles of all pixels reduces noise and thus enables more robust peak detection. The results are presented in Fig. \ref{fig:CrossingAngle_Diameter}A. For both 32 ($\Delta \phi = 11.25^\circ$) and 64 ($\Delta \phi = 5.6^\circ$) illumination angles, the reconstructed crossing angles agree well with the theoretical values. The smaller errors for 64 illumination angles can be attributed to the
higher angular resolution.
The results show that SLI can reliably reconstruct directional information of intersecting structures.

In addition to directional information, the scattered signal also contains information about the geometry of the structure due to the scattering form factor described in equation \ref{eq:form_factor}. Using eq. \ref{eq:diameter_rec}, the structure diameter can be determined from FWHM of the peaks. To do this, the manually masked rods are analysed individually. For each pixel, the mean FWHM of all associated peaks is determined as a first step,
and subsequently the associated diameter calculated. Finally,
the diameters of all pixels in a rod are then averaged. The results are shown in Figure \ref{fig:CrossingAngle_Diameter}B1-B3.
Whilst diameters of the correct order of magnitude are reconstructed for smaller rods (samples 2 and 3, see Fig. \ref{fig:CrossingAngle_Diameter}B2/B3), the diameter of the larger structure (sample 1 in Fig. \ref{fig:CrossingAngle_Diameter}B1) is significantly underestimated. One possible cause is that the scattering is influenced more strongly by the edge regions of the glass rod than by its interior. As the model used assumes that the entire rod diameter scatters homogeneously, this could lead to a systematic underestimation of the reconstructed diameter.

Furthermore, asymmetries in the intensity peaks were observed. In the theoretical model, the ideal cylinder has no preferred illumination direction, so both peaks are expected to have equal intensity. Such asymmetries may influence the peak width and consequently the reconstructed diameter. A possible explanation is scattering artifacts originating from neighbouring structures. Reducing these artifacts could therefore improve the robustness of the diameter reconstruction, for example by matching the refractive index via submersion of the samples in a water-glycerol solution.

\subsection{Tilted glass phantoms}

To evaluate the inclination reconstruction described by Eq.~\ref{eq:inclination_peak_separation}, 3D-printed glass phantoms with a nominal crossing angle of $45^\circ$ were measured in the tilted-sample geometry shown in Fig.~\ref{fig:setup}B. The samples were tilted in steps of $\Delta\epsilon = 5^\circ$ over the interval $\epsilon = [0^\circ,70^\circ]$. The two glass strands were oriented at $\phi_1 = 105^\circ$ and $\phi_2 = 150^\circ$, respectively. Since the sample was tilted around an in-plane axis with orientation $\psi = 90^\circ$, the expected apparent in-plane orientation and inclination do not correspond directly to the mechanical tilt angle $\epsilon$.

The expected orientation of a strand with initial in-plane orientation $\phi_k$ was computed by rotating its direction vector around the tilt axis. The projected in-plane orientation is given by
\begin{equation*}
\phi_{k,\mathrm{exp}}
=
\operatorname{atan2}
\left(
v_{k,y},
v_{k,x}
\right)
\bmod \pi ,
\end{equation*}
with
\begin{align*}
v_{k,x}
&=
\cos\phi_k\cos\epsilon
+
\cos\psi\cos(\phi_k-\psi)
\left(1-\cos\epsilon\right), \\
v_{k,y}
&=
\sin\phi_k\cos\epsilon
+
\sin\psi\cos(\phi_k-\psi)
\left(1-\cos\epsilon\right).
\end{align*}
The corresponding expected polar inclination parameter is
\begin{equation*}
\theta_{k,\mathrm{exp}}
=
\arccos
\left(
\left|
\sin\epsilon,\sin(\psi-\phi_k)
\right|
\right) \ .
\end{equation*}
Here, $\theta_{k,\mathrm{exp}} = 90^\circ$ corresponds to a strand lying in the imaging plane. The complementary out-of-plane angle is therefore $90^\circ-\theta_{k,\mathrm{exp}}$.

The sample region was segmented from the camera image using an intensity- and contrast-based Otsu thresholding procedure. Because large tilt angles introduced optical distortions and intensity inhomogeneities, an optional Gaussian-mixture-model correction was applied to refine the sample mask.

The peak-pair attribution algorithm described in Section~\ref{sec:algorithm} was applied pixelwise to the normalized angular profiles within the segmented sample regions. The profiles were smoothed with a Gaussian kernel of $\sigma = 1.6$ angular bins. Peaks were detected using a relative height threshold of $8\%$, a minimum peak width of one bin, and a minimum peak separation of $45^\circ$. Shared peaks were penalized with a weight of $8$. In addition to the direct peak-pair attribution, each attributed peak pair was used to initialize a nonlinear fit of the $\mathrm{sinc}^2$ model derived from Eq.~\ref{eq:form_factor}. The angular fit parameters were left unconstrained, whereas the reconstructed diameter was restricted to the physically relevant interval $D \in [0.5,10] \mu\mathrm{m}$.

\begin{figure}
\centering
\includegraphics[width=0.95\linewidth]{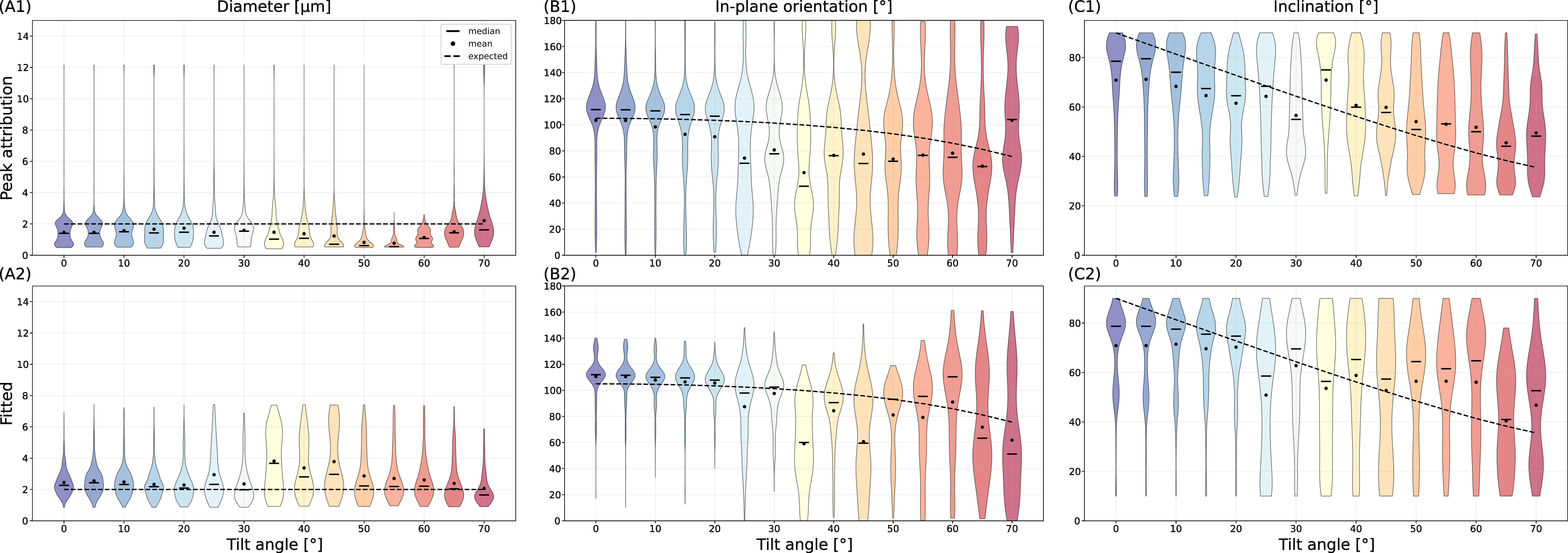}
\caption{Reconstruction of physical parameters from measurements of a tilted glass sample using the geometry shown in Fig.~\ref{fig:setup}B. The reconstructed quantities are the cylinder diameter (A), in-plane orientation (B), and inclination angle (C). Violin plots show the distribution of reconstructed values for each sample tilt angle, sampled in steps of $5^\circ$. The black dashed lines indicate the expected values. Panels A--C1 show the results obtained by peak-pair attribution, whereas panels A--C2 show the corresponding fit-based reconstruction results. The median and mean are marked by a solid vertical line and a dot, respectively.}
\label{fig:Tilted_results}
\end{figure}

Figure~\ref{fig:Tilted_results} shows the reconstructed diameter, in-plane orientation, and inclination for the tilted glass phantom. The direct peak-pair attribution results are shown in panels A--C1, whereas the fit-based results are shown in panels A--C2.

The peak-pair attribution estimates diameters mainly in the range of $0.5$--$3 \mu\mathrm{m}$, with no clear systematic dependence on the sample tilt angle. The reconstructed diameter is slightly underestimated, with typical values around $D_{\mathrm{rec}} \approx 1.5 \mu\mathrm{m}$ compared with the nominal value of $D = 2 \mu\mathrm{m}$. The subsequent fit increases the estimated diameter to approximately $D_{\mathrm{rec}} \approx 2.25 \mu\mathrm{m}$, thereby improving the agreement with the expected value for most tilt angles. Deviations remain most pronounced around $\epsilon = 35^\circ$--$45^\circ$.

The in-plane orientation reconstructed by peak-pair attribution agrees well with the expected orientation for tilt angles up to approximately $20^\circ$ (Fig.~\ref{fig:Tilted_results}B1). In this range, the median values remain close to the expected values, with deviations of approximately $\Delta\phi \approx 5^\circ$, which is below the angular sampling interval of $11.25^\circ$. At larger tilt angles, the distributions broaden substantially and partially extend over the full angular range. Although the reconstructed orientations still follow the expected trend qualitatively, they are generally underestimated, with the largest deviations occurring between $\epsilon = 25^\circ$ and $50^\circ$. For tilt angles between $55^\circ$ and $65^\circ$, the reconstructed orientations again approach the expected range.

The fit-based reconstruction considerably narrows the orientation distributions (Fig.~\ref{fig:Tilted_results}B2). Most reconstructed orientations are then centered close to the expected values. Remaining deviations are mainly observed at tilt angles of $35^\circ$, $45^\circ$, and $60^\circ$.

The inclination reconstructed by peak-pair attribution follows the overall trend of the expected inclination curve (Fig.~\ref{fig:Tilted_results}C1). For small tilt angles up to approximately $20^\circ$, the distributions are centered close to the expected values. At larger tilt angles, the distributions broaden, and the median and mean values show deviations of approximately $10^\circ$ from the expected inclination. The fit-based reconstruction reduces the spread of the inclination distributions (Fig.~\ref{fig:Tilted_results}C2), particularly for tilt angles between $10^\circ$ and $35^\circ$, where the reconstructed values approach the expected curve more closely. For larger tilt angles, especially between $50^\circ$ and $60^\circ$, the inclination is systematically underestimated. At $\epsilon = 70^\circ$, the inclination is not reliably reconstructed, which is expected because this configuration lies outside the effective detection range of the setup.

Despite deviations from the expected values, the newly introduced reconstruction relations for inclination and diameter, given in Eqs.~\ref{eq:inclination_peak_separation} and ~\ref{eq:diameter_rec}, enable an approximate estimation of both parameters from the measured SLI profiles. The observed deviations are likely caused by a combination of model limitations, finite angular sampling, peak-attribution errors, and the scattering artifacts described in Section~\ref{sec:characterization}. Nevertheless, the reconstructed values follow the expected trends sufficiently well to justify applying the proposed evaluation approach to tissue measurements.

\subsection{Application to human brain tissue}

\begin{figure}
\centering
\includegraphics[width=0.95\linewidth]{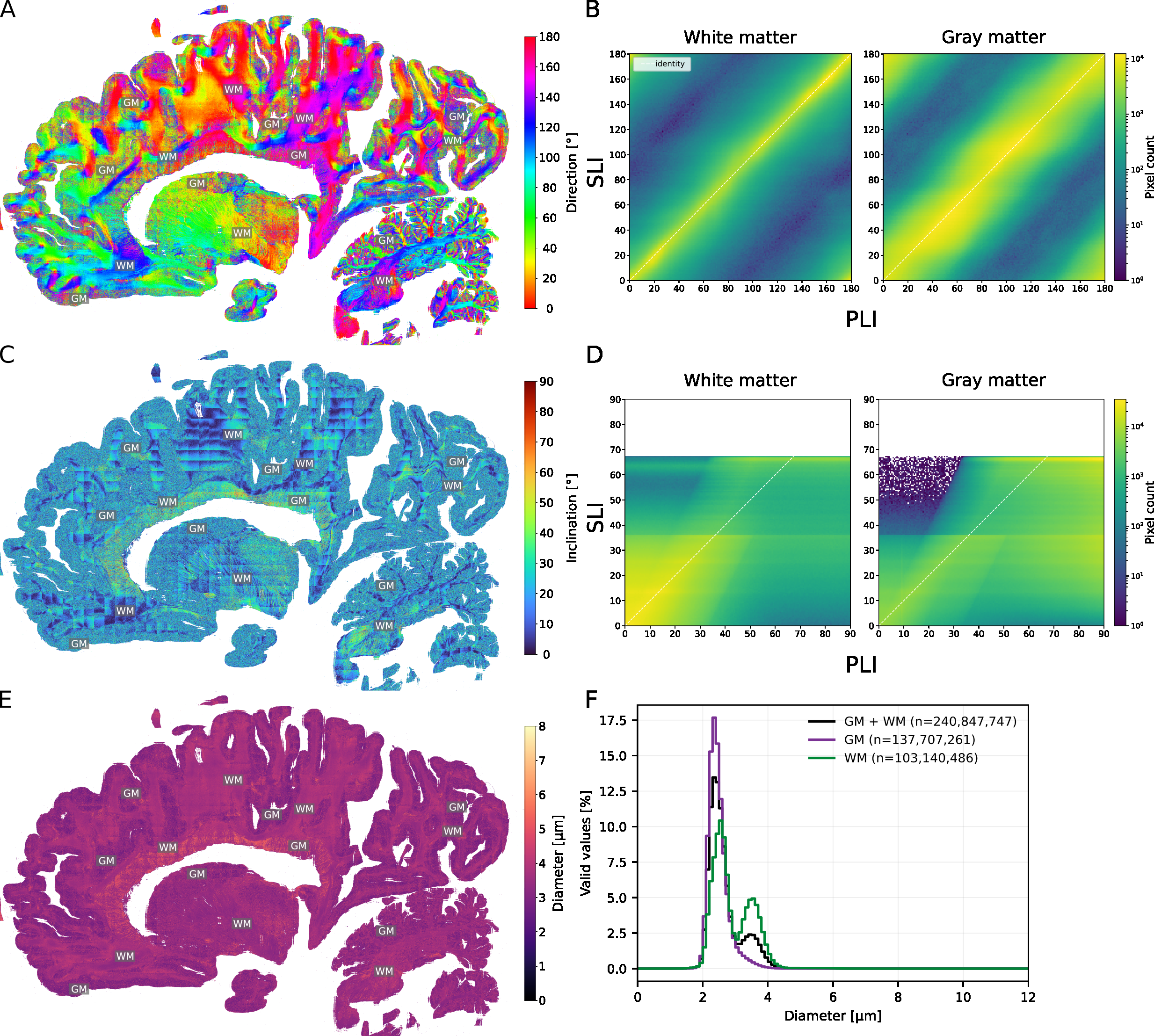}
\caption{3D SLI applied to the tissue section shown in Fig.~\ref{fig:setup}C. A: Reconstructed in-plane orientation map. B: 2D histogram of SLI orientation versus PLI orientation for white matter (left) and gray matter (right). C: Reconstructed inclination map. D: 2D histogram of SLI inclination versus PLI inclination for white matter (left) and gray matter (right). E: Reconstructed diameter map. F: Histogram of reconstructed diameter values for the full tissue section (black), gray matter (purple), and white matter (green). Several gray matter (GM) and white matter (WM) are marked in insets A, C \& E.}
\label{fig:tissue_results}
\end{figure}

Figure~\ref{fig:tissue_results} presents the peak-pair attribution results for the full tissue section shown in Fig.~\ref{fig:setup}C. 
The sample is a sagittal section of the human brain and can be roughly divided into two main constituents: the cortical tissue that contains predominantly neurons and supporting cell types, which is located at the surface of the brain, called gray matter (GM) and the nerve fiber projections that form thick fiber bundles and connect different cortical structures, called white matter (WM). A whole-brain section was selected to include a broad range of organizational structures in both GM and WM without going into anatomical detail and to demonstrate the scalability of the proposed evaluation method.

The following parameters were used for the peak-pair attribution algorithm. The profiles were smoothed with a Gaussian kernel of $\sigma = 0.5$ angular bins. Peaks were detected using a relative height threshold of $10\%$, a minimum peak width of three bins, and a minimum peak separation of $18^\circ$. Shared peaks were penalized with a weight of $0.35$. In addition, a Gaussian mixture model, as previously applied in \cite{Scheidt:26_Hadamard}, was applied to the average scattering intensity to segment white matter, gray matter, and background.

Figure~\ref{fig:tissue_results}A shows the resulting in-plane orientation map, encoded in the HSV color space. The white matter exhibits spatially homogeneous orientations that agree well with the anatomical orientation of the major white matter pathways. In contrast, the gray matter, particularly the cortex, appears noisier and less sharply resolved. To evaluate the performance of the peak-pair attribution algorithm, the reconstructed SLI orientation was compared pixelwise with the in-plane fiber orientation obtained by 3D-PLI. Figure~\ref{fig:tissue_results}B shows this comparison as 2D histograms for white matter (left) and gray matter (right). Perfect agreement would result in a straight line along the identity diagonal, indicated by the white dashed line.\\
In white matter, the SLI orientation shows an almost ideal correlation with the 3D-PLI orientation, with only small deviations around the identity diagonal. Two additional clusters occur at $\phi_{\mathrm{SLI}} = 180^\circ$ and $\phi_{\mathrm{PLI}} = 0^\circ$, and vice versa. These clusters result from the $180^\circ$ periodicity of the axial orientation representation. In gray matter, the histogram shows a similar overall correlation, but the distribution is considerably broader. This is likely caused by the high sensitivity of 3D-PLI to fiber orientations in cortical gray matter \cite{PLIpaper}, whereas SLI is more strongly affected by the lower axonal density and weaker anisotropic scattering signal in this region \cite{SLIpaper}.

Figure~\ref{fig:tissue_results}C shows the reconstructed inclination map. Visually, the tissue can again be separated into white matter with comparatively smooth inclination values and gray matter with noisier, more spatially heterogeneous inclination estimates. Similar to the orientation map, the reconstructed white matter inclinations are consistent with the expected anatomical fiber architecture. Figure~\ref{fig:tissue_results}D compares the SLI inclination with the PLI inclination on a pixelwise basis using 2D histograms for white matter (left) and gray matter (right). The SLI inclination is limited to approximately $67.5^\circ$, since larger inclinations cannot be reconstructed for the given polar illumination angle.

Compared with the in-plane orientation, the inclination shows a weaker correlation between SLI and 3D-PLI. In the white matter, the agreement is best for inclinations up to approximately $35^\circ$, where the distribution follows the identity line closely. However, PLI inclination values close to zero are partly assigned SLI inclinations of up to $20^\circ$. In addition, a horizontal accumulation of values occurs around $\theta_{\mathrm{SLI}} = 35^\circ$, indicating that fewer pixels are assigned to higher SLI inclinations. This behavior is likely related to the peak-separation constraints used in the peak-pair attribution algorithm, which favor peak pairs with larger angular separations.

In gray matter, high PLI inclination values are mapped to a broad and comparatively homogeneous range of SLI inclination values. This discrepancy may partly result from the reduced reliability of PLI inclination reconstruction in regions with low retardation values \cite{PLIpaper}. Nevertheless, the gray matter distribution still follows the identity line within a broader margin of deviation.

Finally, Fig.~\ref{fig:tissue_results}E shows the reconstructed diameter map. A clear visual contrast between gray and white matter is observed. White matter exhibits larger reconstructed diameters than gray matter and appears more spatially homogeneous and structured. Figure~\ref{fig:tissue_results}F shows the corresponding diameter histograms for the full tissue section (black), gray matter (purple), and white matter (green). The gray matter distribution has a single dominant peak at approximately $D \approx 2.5\mu\mathrm{m}$. In contrast, the white matter distribution shows two peaks, one near $D \approx 2.5\mu\mathrm{m}$ and a second near $D \approx 3.8\mu\mathrm{m}$. This is consistent with the expected anatomical differences between gray and white matter composition \cite{axon_diameter}.

\section{Conclusion}

The results demonstrate that the proposed reconstruction approach can estimate in-plane orientation, inclination, and effective fiber diameter from angular SLI profiles, although several limitations remain. Measurements of tilted glass phantoms showed that the Rayleigh--Gans-based model captures the expected trends of the reconstructed parameters, but deviations from the nominal values were observed. These deviations are likely caused by a combination of model limitations, finite angular sampling, peak-attribution errors, and sample-specific artifacts. In particular, the two-photon-lithography fabrication process can introduce impurities and small deformations during shrinkage and cooling. Furthermore, the glass phantoms were measured in air, such that scattering at the glass-air interface was not suppressed. This differs from tissue measurements, where refractive-index matching by the mounting medium reduces surface-related scattering artifacts.

Future phantom measurements will therefore be performed in a refractive-index-matching medium to determine whether edge artifacts can be reduced and whether the agreement with the expected diameter and inclination values improves are therefore necessary. In addition, scanning electron microscopy and beam-profiling measurements will be acquired to better characterize the phantom geometry and optical quality. Since these measurements are destructive, they will be performed after completing further optical measurements.

The applicability of the model is supported by the measurements in human brain tissue. The reconstructed in-plane SLI orientation agrees well with the corresponding 3D-PLI orientation, particularly in white matter. Remaining deviations are currently under investigation and may be region-specific, for example in fiber-crossing regions, where the PLI signal represents an averaged orientation of multiple fiber populations. In gray matter, the SLI signal is less sensitive than the PLI signal because of weaker anisotropic scattering contributions and a generally lower signal strength. Nevertheless, a clear correlation between the SLI- and PLI-derived in-plane orientations is observed.
The reconstructed diameter map shows larger effective diameters in white matter than in gray matter, which is consistent with the expected tissue composition. However, the absolute accuracy of the diameter reconstruction requires further validation using samples with independently measurable fiber diameters, such as spinal cord sections.

The inclination reconstruction agrees well with 3D-PLI for small inclination values, especially in white matter. Larger deviations occur for high inclinations, where the PLI data show a stronger representation of steeply inclined fibers than the SLI reconstruction. These deviations may arise from differences in the physical contrast mechanisms of SLI and 3D-PLI, from reduced sensitivity of the present SLI geometry at high inclinations, or from uncertainties in the peak-pair attribution. Further validation is therefore required, but the present results indicate that the proposed SLI-based reconstruction provides a promising route toward estimating three-dimensional fiber architecture and effective fiber diameter in unstained human brain tissue.

\section*{Funding} Work partially funded by AIDAS - AI, Data Analytics and Scalable Simulation - which is a Joint Virtual Laboratory gathering the Forschungszentrum Jülich (FZJ) and the French Alternative Energies and Atomic Energy Commission (CEA); The European Union’s Horizon Europe Program under the Specific Grant Agreement No. 101147319 (EBRAINS 2.0 Project); And the German Research Foundation (DFG) Project no. 498596755. Computing time was granted through VSR Computing Time Projects on the supercomputer JURECA at Jülich Supercomputing Centre (JSC), Germany.

\section*{Acknowledgment}  Special thanks to the lab team of INM-1 (Forschungszentrum Jülich GmbH, Germany) for preparing the brain sample, with particular appreciation to Anna Steffens, Phillip Schlömer, and Markus Cremer. 
Generative AI was used for spelling and grammar correction.

\bibliographystyle{elsarticle-harv} 
\bibliography{bibliography}

\end{document}